\documentclass{caosp306}

\usepackage{graphicx}

\usepackage{natbib}
\articleNo{000}
\pubyear{2019}
\volume{00}
\volnumber{0}
\firstpage{1}
\received{December 16, 2018}
\accepted{quickly}

\def\BibTeX{{\rm B\kern-.05em{\sc i\kern-.025em b}\kern-.08em
             T\kern-.1667em\lower.7ex\hbox{E}\kern-.125emX}}

\begin{document}

\hauthor{T.\,Zwitter}

\title{Galactic astronomy and small telescopes}

\author{
        Toma\v{z}\, Zwitter       
}
\institute{University of Ljubljana, Faculty of Mathematics and Physics, \\ Jadranska 19, 1000 Ljubljana, Slovenia \email{tomaz.zwitter@fmf.uni-lj.si}
          }

\date{tbd}
\maketitle

\begin{abstract}
The second data release of ESA's Gaia satellite (Gaia DR2) revolutionised astronomy by providing accurate distances, proper motions, apparent magnitudes, and in many cases temperatures and radial velocities for an unprecedented number of stars. These new results, which are freely available, need to be considered in virtually any stellar research project, as they provide crucial information on luminosity, position, motion, orbit, and colours of observed targets. Ground-based spectroscopic surveys, like RAVE, Gaia-ESO, Apogee, LAMOST, and GALAH, are adding more measurements of radial velocities and, most importantly, chemistry of stellar atmospheres, including abundances of individual elements. We briefly describe the new information trove, together with some warnings against blind-folded use. 

Even though it may seem that Gaia is already providing any information that could be collected by small telescopes, the opposite is true. In particular, we discuss a possible reach of a ground-based photometric survey using a custom filter set. We demonstrate that it can provide valuable information on chemistry of observed stars, which is not provided by Gaia or other sky surveys. A survey conducted with a small telescope has the potential to measure both the metallicity and alpha enhancement at a $\sim 0.1$~dex level for a large fraction of Gaia targets, a valuable goal for galactic archaeology. 
\end{abstract}

\section{Introduction}
\label{intro}
Our Galaxy is a typical barred spiral galaxy. So understanding its structure, evolution, and origin is an important endeavour and the goal of galactic archaeology \citep{freeman02}. Many questions can be answered only by studying our Galaxy, as even detection of individual stars in nearby galaxies can be a difficult task. On the other hand we are located inside our Galaxy, so its more than $10^{11}$ stars are distributed all over the sky and are in many cases obscured by interstellar dust clouds. Knowledge of distances and 3-dimensional motions is crucial to derive luminosities and orbits of observed stars. Orbits are important, as they allow identification of a given star as a member of a particular galactic component, i.e.\ the thin or thick disks, bulge or halo. Stars in the halo are moving in a low density environment, so they get perturbed only twice per Galactic orbit during their passages through the Galactic plane. Their kinematic properties, like the total energy, total angular momentum and the size of its z-component maintain a nearly constant value. As a consequence, present orbits of stars in the halo can have their origins traced to individual dwarf galaxies which were cannibalised by our Galaxy some 12 billion years ago \citep{helmi00}. Unfortunately this type of kinematic reasoning does not work in a higher-density environment of a galactic disc. Studies of origin of individual disk stars should therefore rely on measurements of chemistry of their atmospheres, which remains constant throughout their lives (except for material dredge-ups late in their lives). So one could hope to identify stellar siblings with an origin in a common star cluster which has now long-gone. This is the idea behind chemical tagging \citep{desilva07,desilva15,kos18}. 

In this contribution we first briefly review recent space and ground-based stellar sky surveys, which now provide a  firm observational base for galactic archaeology. We urge the users to use the new and freely available datasets in their daily research activities. Finally we connect these datasets with the reach of surveys using small telescopes, in particular we discuss how stellar chemistry can be studied through a photometric survey which uses a customised filter set.

\section{Gaia satellite of ESA}
\label{gaia}

April 25 2018 is an important date, as from then on astronomers have access to the second data release of the satellite Gaia of the European Space Agency \citep[Gaia DR2, ][]{brown18}. The release vastly expanded the available number (Table \ref{table1}) and accuracy (Table \ref{table2}) of astrometric, but also photometric and radial velocity measurements. Gaia DR2 profoundly changed every area of astronomy: parallaxes, for example, are now available for 1.3 billion stars at a typical accuracy of 0.03~mas at the bright end, which can be compared to already very impressive but much smaller and less accurate parallaxes for 2 million stars at a 0.3~mas level, published in Gaia's first data release \citep{brown16}. 

\begin{table}[t]
\small
\begin{center}
\caption{Number of stars in the Gaia DR2 with a cited type of information \citep{brown18}.}
\label{table1}
\begin{tabular}{lr}
\hline\hline
Type of information & Number of stars\\\hline
Position and brightness on the sky & 1,692,919,135\\
Red colour photometry                    & 1,383,551,713\\
Blue colour photometry                  & 1,381,964,755\\
Parallax and proper motion            & 1,331,909,727\\
Surface temperature                      &    161,497,595\\
Reddening along the line of sight    &      87,733,672\\
Radius and luminosity                     &      76,956,778\\
Radial velocity                                 &        7,224,631\\
Variable sources                             &          550,737\\
Orbits of Solar system objects        &            14,099\\
\hline\hline
\end{tabular}
\end{center}
\end{table}

\begin{table}[t]
\small
\begin{center}
\caption{Typical uncertainties of Gaia DR2 \citep{brown18}. $G$ magnitude is a white light measurement by Gaia, while $G_{BP}$, $G_{RP}$ and $G_{RVS}$ are integral measurements in its blue, red and Gaia-RVS bands. mas stands for milli-arc-second.}
\label{table2}
\begin{tabular}{lr}
\hline\hline
Data product or source type & Typical uncertainty\\\hline
Five-parameter astrometry (position \&\ parallax) & 0.02-0.04 mas at $G<15$\\
 & 0.1 mas at $G=17$\\
 & 0.7 mas at $G=20$\\
 & 2 mas at $G=21$\smallskip\\
Five-parameter astrometry (proper motion) & 0.07 mas yr$^{-1}$ at $G<15$\\
 & 0.2 mas yr$^{-1}$ at $G=17$\\ 
 & 1.2 mas yr$^{-1}$ at $G=20$\\ 
 & 3 mas yr$^{-1}$ at $G=21$\smallskip\\ 
Mean G-band photometry & 0.3 mmag at $G<13$\\
 & 2 mmag at $G=17$\\
 & 10 mmag at $G=20$\smallskip\\
Mean $G_{\rm BP}$ and $G_{\rm RP}$ band photometry & 2 mmag at $G<13$\\
 & 10 mmag at $G=17$\\
 & 200 mmag at $G=20$\smallskip\\
Median radial velocity over 22 months & 0.3 km\,s$^{-1}$ at $G_{\rm RVS}<8$\\
 & 0.6 km\,s$^{-1}$ at $G_{\rm RVS}=10$\\
 & 1.8 km\,s$^{-1}$ at $G_{\rm RVS}=11.75$\smallskip\\
 Systematic radial velocity errors & $<0.1$~km\,s$^{-1}$ at $G_{\rm RVS}<9$\\
  &0.5~km\,s$^{-1}$ at $G_{\rm RVS}<11.75$\\
\hline\hline
\end{tabular}
\end{center}
\end{table}

New information should benefit virtually any stellar research project. In many cases it provides a good handle on absolute magnitude, colour, spatial position and orbit of the observed object(s), a completely novel situation in astronomy where such information has been largely limited to members of star clusters or other spatially confined groups where a suitable standard candle could be found. Finally the field stars are joining the game of disentangling detailed physics and Galactic orbits. 

The data trove of Gaia should be, however, used responsibly. First, Gaia is not measuring distances but parallaxes. Relation between the two can be non-trivial, as thoroughly discussed by \citet{bailerjones18}. Some of the reported parallaxes are negative which {\it is} a correct statistical data treatment. In many cases this puts a firm lower limit on the true distance of the observed object, which can be a valuable piece of information. Some parallaxes are very large, 59 of them have a value larger than the parallax of Proxima Centauri. Still, this does not mean they are closer, but that the reported value is a result of an automated procedure which can get confused by spurious alignment with sources which are resolved by Gaia only in some observations and not in others. It is important to note that there is no universal recipe to mitigate such cases, but that the users can perform custom tests which will effectively resolve such issues on parallaxes and/or proper motions. A useful example is discussed in figure C4 of \citet{lindegren18}.

\begin{table}[t]
\small
\begin{center}
\caption{Typical errors on distance, proper motion and velocity along the sky plane for a star at a distance of 1~kpc if 
effects of interstellar reddening can be neglected. Errors on velocity along the sky plane depend on actual velocity 
of the object: $\sigma(V_{\mu})_1$ lists the error if the object is at rest and the error is driven by the proper motion uncertainty; and 
$\sigma(V_{\mu})_2$ is for an object with a velocity of 50~km~s$^{-1}$ along the sky plane, where the error is driven by the 
distance error.}
\label{table3}
\begin{tabular}{llrr}
\hline\hline
Parameter&Symbol&Solar type star& Red clump star\\\hline
Absolute $V$ magnitude& $M_V$   & 4.83 & 0.5\\
$V-I_{\rm C}$ colour                & $V - I_{\rm C}$& 0.70& 1.06\\
Apparent $G$ magnitude& $G$      &14.5      & 10.1\\
Parallax error                   & $\sigma_{\tilde{\omega}}$ &30 $\mu$as & 40 $\mu$as\\ 
Distance error                  & $\sigma_{distance}$  & 3\%  & 4\% \\
Proper motion error         & $\sigma_\mu$ & 17 $\mu$as~yr$^{-1}$ & 22 $\mu$as~yr$^{-1}$\\
Velocity error along the sky plane&$\sigma (V_{\mu})_1$ &  0.08~km~s$^{-1}$ & 0.11~km~s$^{-1}$\\
Velocity error along the sky plane&$\sigma (V_{\mu})_2$ &  1.5~km~s$^{-1}$ & 2~km~s$^{-1}$\\
\hline\hline
\end{tabular}
\end{center}
\end{table}

Energy and angular momentum of a stellar Galactic orbit cannot be computed without knowledge of the velocity vector. While proper motion and parallax measurements provide accurate values of velocity components along the sky plane, radial velocity can be measured only by spectroscopy, except for nearby stars where perspective acceleration could be used \citep{prusti16,debruijne12}. Gaia mission measures radial velocities with an onboard spectrograph and Gaia DR2 contains an unprecedented set of radial velocities for 7,2 million objects with a typical error of $\sim 1$~km~s$^{-1}$ \citep{cropper18,katz18,sartoretti18}. This error can be compared to the error budget on velocities along the sky plane.
Table \ref{table3} illustrates the error budget for a Solar type and for a red clump star, both at a distance of 1 kpc, if interstellar extinction can be neglected. The next to last line shows that for stars at rest their velocities along the sky plane are determined to $\sim 0.1$~km~s$^{-1}$, while the last line demonstrates that for fast moving stars the uncertainties on their distance increase the errors on their velocities along the sky plane to more than 1~km~s$^{-1}$. So at least for slowly moving stars and for the ones in clusters it is desirable that the radial velocity component has an error similar to errors of velocities along the sky plane, i.e.\ $\sim 0.1$~km~s$^{-1}$. Note also that any studies of dynamics {\it inside} star clusters greatly benefit from accurate velocities, as  escape velocity from a cluster seldom supersedes a few km~s$^{-1}$.

\section{Ground based spectroscopic surveys}
\label{ground}

Ground-based spectroscopic surveys designed to complement Gaia include RAVE, Gaia-ESO, Apogee, GALAH, and LAMOST. All but the last one measure radial velocities at a level better than or comparable to Gaia, but for a much smaller number of stars. Internal precision of Apogee is better than 0.1~km~s$^{-1}$ \citep{nidever15}, and the accuracy is $\sim 0.3$~km~s$^{-1}$ \citep{anguiano18}. GALAH survey currently observed more than half-a-million stars with a typical accuracy of their radial velocity of $\sim 0.15$~km~s$^{-1}$ \citep{zwitter18}, as it includes also effects of convective blueshift and gravitational redshift of light emerging from a stellar atmosphere. Note that the GALAH survey published also medians of observed spectra that are nearly noiseless, as they are obtained from a large number of observed spectra belonging to the same bin in stellar parameter space with a width of 50 K in temperature, 0.2 dex in gravity, and 0.1 dex in metallicity. Publicly released 1181 median spectra have a resolving power of 28 000 and trace the well-populated stellar types with metallicities between $-0.6$ and $+0.3$. 

All of the mentioned ground-based spectroscopic surveys determine values of stellar parameters and in many cases also abundances of individual elements. The GALAH survey recently published what is currently the largest set of abundances of up to 23 chemical elements, which now includes 342,682 stars \citep{buder18}. This chemical information is crucial in complementing the excellent astrometric and photometric information from Gaia.

\section{Proposal for a dedicated photometric survey}

Gaia can judge on chemical composition of stars from two sources of information. It can use its radial velocity spectrograph (RVS)  which collects several tens of spectra for stars brighter than $V \sim 15$. By adding up all spectra of a given star the combined spectrum yields radial velocity and -- for bright-enough sources ($V\la 12.5$) -- also metallicity. Experience from the RAVE survey,  which observed in the same wavelength range as Gaia RVS, shows that metallicity is driven by strong lines of the Calcium triplet, so that it is very difficult to separate iron abundance ([Fe$/$H]) from alpha enhancement ([$\alpha/$Fe]) for relatively faint stars with noisy spectra \citep{zwitter08,kunder17}. The second source of chemistry for Gaia is star's spectral energy distribution which is sampled by low dispersion BP and RP spectra. Wavelength span of an effective resolution element (defined as 
76\%\ energy extent of the along-scan line spread function) varies from 9.1~nm at 330~nm to 24.5~nm at 640~nm and to 61.5~nm at 1050~nm \citep{prusti16}. Assuming the star is observed 100 times during the mission its total exposure time for the BP and RP instruments equals 442~seconds, each. These low dispersion spectra can be used to determine values of astrophysical parameters for fainter stars than with the RVS instrument. As explained in \citet{bailerjones11} and \citet{prusti16} for FGKM stars at $G = 15$ with less than two magnitudes extinction, effective temperature can be estimated to $\sim 150$~K, extinction to $\sim 0.1$ mag, surface gravity to $\sim 0.3$~dex, and metallicity [Fe/H] to $\sim 0.2$~dex. Ground-based spectroscopic surveys (RAVE, Apogee, Gaia-ESO, LAMOST, GALAH) have been useful in providing metallicities and in some cases element abundances, but cumulatively they now include less than 5 million stars, a small fraction of 160 million stars brighter than $G = 17$ with effective temperatures published by Gaia DR2. Here we propose a way to extend results on chemistry to fainter stars than doable by Gaia and to obtain iron abundances ([Fe$/$H]) and alpha-enhancement ($\alpha/$Fe] separately. This is vital, as iron and alpha element abundances form a basis for stellar population studies \citep[e.g.\ ][]{krumholz18}, and present a useful chemical clock \citep[e.g.\ ][]{freeman02}. 

Star's chemistry can be inferred by measuring its brightness in 3 specific photometric filter bands centred on and off the prominent metallic lines. We use three readily available industrial photometric filters, which have however never been used in astronomy. Band A ($394 \pm 5$~nm) is centred on the Ca~II H\&K lines, so it is sensitive to abundance of alpha elements ([$\alpha/$Fe]), band B ($430 \pm 5$~nm) contains many strong Fe~I lines, so it is sensitive to metallicity ([Fe$/$H]), while band C ($450 \pm 5$~nm) lacks any strong metallic lines. Bands have a width of 10~nm, thus moderately sized telescopes can be used. These are hard coated OD 4 10~nm bandpass filters in the Edmund Optics catalogue (items \#65-192, \#65-198, and \#65-201). Their wide availability means that results obtained by different instruments can complement each other. 

\begin{figure}
\centerline{\includegraphics[width=0.64\textwidth,height=0.9\textwidth,angle=270,clip=]{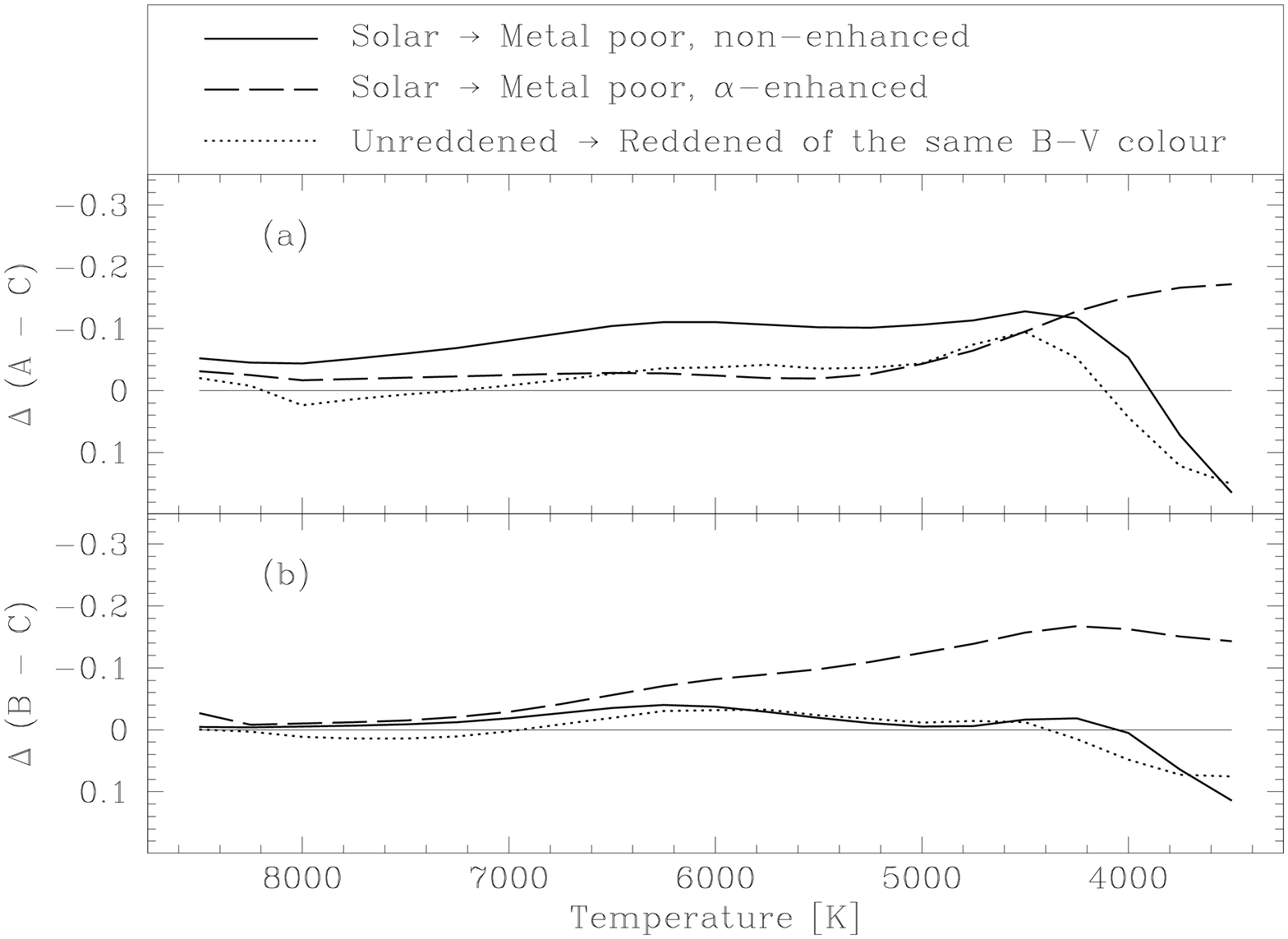}}
\caption{Change in the $A-C$ spectral index (panel a) and $B-C$ spectral index (panel b) as a function of effective temperature for main sequence stars. Three types of changes are plotted: (i) solid lines: observation of a metal poor non-enhanced star ($[{\rm Fe}/{\rm H}]=-0.4$, $[\alpha/{\rm Fe}]=0.0$) instead of a Solar type one ($[{\rm Fe}/{\rm H}]=0.0$, $[\alpha/{\rm Fe}]=0.0$) with the same temperature; (ii) dashed lines: observation of a metal poor $\alpha$-enhanced star ($[{\rm Fe}/{\rm H}]=-0.4$, $[\alpha/{\rm Fe}]=+0.4$) instead of a Solar type one ($[{\rm Fe}/{\rm H}]=0.0$, $[\alpha/{\rm Fe}]=0.0$) with the same temperature; (iii) dotted lines: observation of two Solar type stars which have the same observed $B-V$ colour, but one is reddened ($A_{\rm V}=0.155$~mag) and the other is not. Horizontal line traces zero values.}
\label{fig1}
\end{figure}

\begin{figure}
\centerline{\includegraphics[width=0.64\textwidth,height=0.9\textwidth,angle=270,clip=]{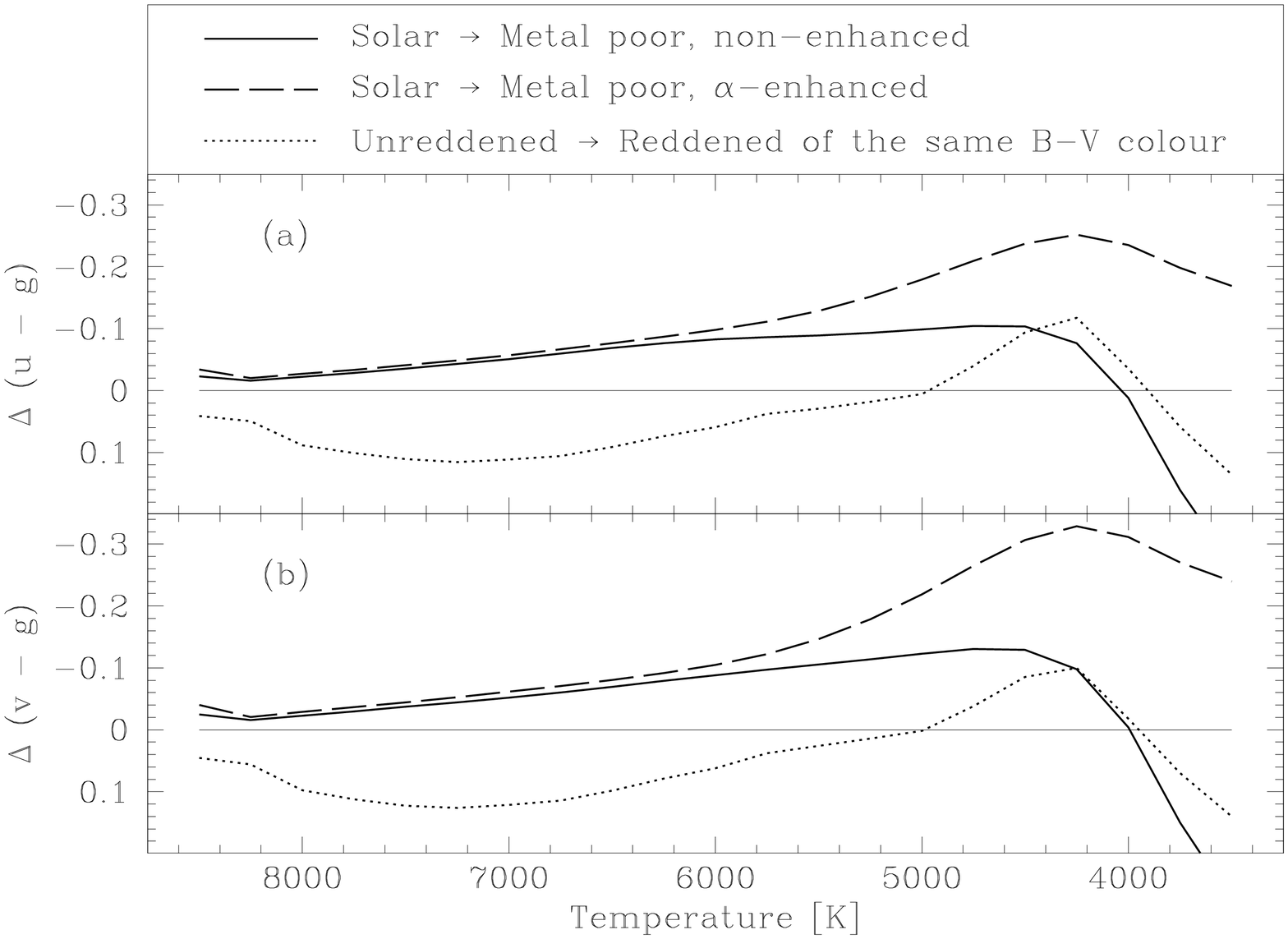}}
\caption{The same as Fig.~\ref{fig1} but for a Sloan $u-g$ index (panel a) and $v-g$ index (panel b). Pairwise behaviour of solid and dashed lines is virtually the same for both panels, so one cannot use the two spectral indices to distinguish changes in metallicity and alpha enhancement. Sloan bands are very blue, so they are sensitive to reddening (dotted lines) which can overshadow changes due to chemistry. }
\label{fig2}
\end{figure}

We build on the fact that luminosity and effective temperature of the source is now known from Gaia, so we can use the colour-colour vs. chemistry relationship holding at this particular position in the luminosity-temperature plane. These relationships can be established by a data driven approach using detailed chemical abundances of benchmark stars measured by ground-based spectroscopic surveys. The GALAH survey \citep{buder18} and other observations by the Hermes spectrograph at the 4-m AAO telescope, which now contain accurate chemical abundances for up to 30 chemical elements and for $\sim 600.000$ stars, can be used, together with the results from RAVE and Gaia-ESO surveys. 

Figure \ref{fig1} illustrates properties of the proposed dataset for main sequence stars \citep[as defined by][]{pecaut13} using Kurucz models of stellar atmospheres \citep{munari05}. AB magnitudes in the three filters are combined into two spectral indices: $A-C$ (top panel) and $B-C$ (bottom panel). For each panel we plot the difference in magnitude if a star with Solar abundances and with a given effective temperature is mistaken for another similar (but not identical) type of object. Solid lines plot the difference if we observe a non-enhanced metal poor star of the same temperature, and dashed lines are for an alpha enhanced metal poor star. Finally, the dotted lines check on the influence of interstellar reddening: a star with Solar abundances with a given temperature is replaced with a reddened main sequence star ($A_V = 0.155$~mag) of Solar composition but with an intrinsically higher effective temperature, so that the observed $B-V$ colours of both stars are the same.  

Absorption lines in the wavelength range of filter $A$ are mostly lines of calcium, an $\alpha$ element. So the $A-C$ index (Fig.~\ref{fig1}a) is sensitive to metal content, in the sense that the flux in filter $A$ increases in the absence of absorption lines. This is true if all elements, including calcium, have a lower abundance (solid line in the top panel). But if calcium abundance is increased to Solar levels in a metal poor $\alpha$-enhanced star ($[{\rm Fe}/{\rm H}]=-0.4$, $[\alpha/{\rm Fe}]=+0.4$) the $A-C$ index stays virtually the same as for a star with Solar abundances. Behaviour of the $B-C$ index (Fig.~\ref{fig1}b) is different. Its values are increased for a metal poor $\alpha$-enhanced star. So a combination of $A-C$ (Fig.~\ref{fig1}a) and $B-C$ (Fig.~\ref{fig1}b) indices allows a {\it separate} judgement of the iron abundance ($[{\rm Fe}/{\rm H}]$) and alpha enhancement ($[\alpha/{\rm Fe}]$) values. 

Values of temperature are known from Gaia, but with a typical error of $\sim 150$~K \citep{prusti16}. Derivation of more accurate values is hampered by unknown values of interstellar extinction which is strongly correlated with temperature. As an example we compare stars with Solar abundances but with two different temperatures: an undereddened star and one which suffers from moderate interstellar reddening ($A_V = 0.155$, $R = 3.1$) and has the same reddened $B-V$ colour as the unreddened star. Typically this implies that a reddened main sequence star is $\sim 160$~K hotter than the unreddened one. Results are shown with a dotted line in Fig.~\ref{fig1}. Clearly the influence of reddening is moderate, except for stars cooler than 4750~K which have a steeply increasing flux in the range of $A$ and $B$ filters. 

Approach outlined here is different from existing photometric surveys (e.g.\ SkyMapper) which use standard Sloan filters in an attempt to determine both the values of stellar parameters and chemistry, with the result that filters are not optimised for chemical analysis. So these surveys find it difficult to separate abundances of iron and alpha group elements, a vital starting point for any stellar population studies. Note also that the forthcoming LSST survey uses a similar un-optimised filter set and will be saturated for most stars observed by Gaia. 

The situation is illustrated in Figure~\ref{fig2} which is equivalent to Figure~\ref{fig1}, except that results for the Sloan indices $u-v$ and $g-v$, as implemented by the SkyMapper survey \citep{bessell11}, are presented. Both indices have almost identical solid and dashed curve shapes, so one can use them to determine general metallicity, but there is little hope to separate alpha enhancement and iron abundance. This agrees with recent results of \citet{casagrande19} who emphasise usefulness of SkyMapper for metallicity but do not mention alpha enhancement determination. Note also that amplitudes of dotted curves in Fig.~\ref{fig2} are much larger than in Fig.~\ref{fig1}, so measurement of metallicity using Sloan $u$ and $g$ filters which are bluer than $A$ and $B$ filters is quite sensitive to damaging effects of interstellar reddening. 

\section{Conclusions}

Gaia DR2 presents a profound change in the way research in astronomy is being done, so it should be used by virtually any astronomer. Still, it is important this information trove is used responsibly, taking into account all the information which was provided along with the data release. Complementary ground-based spectroscopic surveys are adding valuable information on chemistry of observed targets. The last months are bursting with activity, the Gaia DR2 paper \citep{brown18} is the most cited paper in the field of astronomy and astrophysics published this year. Among the many significant results one could point to a disagreement between the value of the Hubble constant as determined from local Cepheids by Gaia and HST and between the value suggested by Planck observations of the early Universe \citep{riess18}, discovery of a perturbation and oscillation of the Galactic disk by a passage of a dwarf galaxy between 300 million and 900 million years ago \citep{antoja18,blandhawthorn18}, formation of the thick disc \citep{helmi18}, rich details of Hertzsprung-Russell diagrams \citep{babusiaux18}, and new studies of stellar clusters which are now relieved from uncertainties regarding their membership (e.g.\ \citeauthor{kos18} 2018, \citeyear{kos18a}). 

Despite the tremendous success Gaia cannot do it all. Very accurate values of radial velocities and abundances of individual chemical elements are largely left to be determined by the ongoing spectroscopic ground-based surveys. An additional possibility which is briefly discussed above is a photometric survey using dedicated filters which should be able to indicate metallicity and alpha enrichment of observed targets. This is achieved through use of dedicated photometric filters which are narrower than resolution elements of Gaia's BP and RP instruments. Such a survey is within reach of a small telescope which could therefore make a very significant contribution to Galactic astronomy. 

\acknowledgements
The author acknowledges financial support of the Slovenian Research Agency (core funding P1-0188 and research project N1-0040).

\end{document}